**Iterative AI-guided optimisation of selective triple-drug combinations for breast cancer**

Oghenejokpeme Orhobor[1,*], Abbi Abdel-Rehim[1,*], Emma Tate[2], Holly X. Smith[2], Elizabeth Bourne[2], Ross J. Collins[2], Larisa N. Soldatova[3], Ross D. King[1,4]

[1]Department of Chemical Engineering and Biotechnology, University of Cambridge, CB3 0AS, U.K.

[2]Arctoris Ltd, Oxford, OX14 4SA, UK.

[3]Department of Computing, Goldsmiths, University of London, SE14 6NW, U.K.

[4]Department of Computer Science and Engineering, Chalmers University, S-412 96 Göteborg, Sweden.

*Corresponding authors: O.O: oo288@cam.ac.uk, A.A: aar52@cam.ac.uk

O.O. and A.A. contributed equally to this study.

**Abstract**

Personalised cancer therapy aims to tailor treatment to individual tumour profiles, yet tumour heterogeneity and adaptive resistance continue to limit clinical efficacy. Drug combinations offer a strategy to overcome resistance by simultaneously targeting multiple pathways, but their rational design is constrained by the vast combinatorial search space and experimental cost.

Here, we present an AI-guided, QSAR-driven iterative optimisation framework that integrates machine learning with automated experimental screening to enable closed-loop discovery of selective multi-drug therapies. Starting from an initial random screen, the system iteratively predicts, tests, and refines three-drug combinations targeting MCF7 breast cancer cells. Incorporation of non-tumorigenic MCF10A cells enables explicit optimisation of tumour-selective efficacy, prioritising regimens that maximise cancer cell killing while sparing healthy cells.

Across successive iterations, the framework rapidly enriched for highly selective, high-efficacy combinations, while maintaining chemical and mechanistic diversity and avoiding convergence on a narrow solution space. By continuously learning from experimental feedback, the approach efficiently navigates millions of combinations to identify a small set of validated, tumour-selective regimens.

These results establish a scalable proof-of-concept for AI-driven, closed-loop optimisation of higher-order drug combinations, demonstrating how iterative integration of computation and experimentation can enable adaptive and potentially personalised therapeutic design in precision oncology.

## Introduction

Despite decades of research, cancer remains a leading cause of mortality worldwide. Advances in understanding its underlying biology and the development of new therapeutic agents have improved outcomes. However, the heterogeneity and adaptability of the disease often lead to treatment failure via resistance mechanisms. One strategy to counteract this issue is the use of drug combinations, which can target multiple pathways simultaneously. Yet, rationally designing effective combinations remains both time- and resource-intensive.

Artificial intelligence (AI) and machine learning have emerged as transformative tools in drug discovery, offering the potential to accelerate development, reduce costs, and provide interpretable recommendations that could guide personalised therapy decisions. Quantitative Structure–Activity Relationship (QSAR) modelling is a long-standing approach for predicting biological activity from molecular features and has been successfully applied in multi-target compound design [1-4].

In contrast, AI applications in personalised medicine are less widespread, though it is widely acknowledged that AI is likely to transform the future of cancer therapy. While genetic profiling is the most common route for patient-specific treatment, only a subset of patients benefit from biomarker-driven therapies, many of which ultimately fail [5,6]. A complementary approach is functional testing, where patient-derived samples are screened against a panel of drugs to identify effective treatments [7-9]. Xenograft models offer biological fidelity but are costly and difficult to scale, limiting their practical utility [10]. Ex vivo cultures derived from patient biopsies provide a more scalable and experimentally tractable alternative while preserving key aspects of tumour biology [11-15]. When combined with iterative AI-guided screening, these systems enable cycles of hypothesis generation, experimental validation, and model refinement under human oversight, forming a workflow that accelerates discovery and improves reproducibility. Screening data from these platforms can then train QSAR models, allowing exploration of broader drug libraries for subsequent validation.

Although QSAR has been applied to predict compound and combination activity in publicly available cell line datasets, most computational studies focus on predicting pairwise drug interactions using static datasets, whereas iterative optimisation of higher-order drug combinations integrated with automated experimentation remains largely unexplored. Notable efforts include studies applying active learning to identify synergistic drug pairs [16, 17] and optimising cocktail formulations rather than combinations themselves [18]. Much of the literature emphasises drug pairs, whereas clinical practice often employs more complex regimens, with patients frequently receiving multiple concurrent therapies. The resulting combinatorial complexity makes exhaustive experimental exploration infeasible, underscoring the need for computational approaches to guide optimisation.

Here, we present a QSAR-guided iterative optimisation workflow to identify selective multi-drug combinations against breast cancer. Using an automated platform and machine learning models, we iteratively predicted, tested, and refined three-drug cocktails, prioritising regimens that maximised tumour cell toxicity while sparing non-tumorigenic cells. This approach addresses gaps in current studies by enabling systematic, scalable optimisation of complex drug combinations beyond pairs and serves as a prototype for AI-driven precision therapy in oncology.

## Results

We applied our QSAR-guided iterative optimisation to identify selective three-drug combinations against MCF7 breast cancer cells while sparing MCF10A. In the initial random screen (Iteration 0), 230 combinations were tested, covering 163 of the 190 available drugs (86.7%) and showing broad chemical diversity, as revealed by principal component analysis (PCA) of MACCS fingerprints (Fig. S1). We first assess the quality of this initial screen, then describe how combination efficacy and selectivity evolved across iterations, and finally examine chemical diversity, individual drug contributions, and mechanistic insights for potent combinations.

### Combination selection over iterations

The optimisation objective was to maximise Δ-viability, the difference in toxicity between MCF7 and MCF10A cells, across Iterations 0–8, resulting in progressively more potent and selective combinations (Fig. 1A). The algorithm rapidly identified potent combinations with good selectivity. A marked increase in average Δ-viability was observed after the first iteration, with steady gains until approximately iteration 5, after which improvements plateaued (Fig. 1B). This plateau likely reflects partial convergence toward a high-performing region of combination space, particularly with respect to maximising MCF7 cytotoxicity, where many combinations were already impacting MCF7 viability substantially.

Notably, subsequent iterations suggest a shift in optimisation behaviour, with continued refinement of selectivity driven by improved tolerance in MCF10A cells. This indicates that, rather than a complete stagnation, the optimisation process transitioned from prioritising tumour cell killing to enhancing differential viability by increasingly sparing non-tumorigenic cells. Additional contributing factors, such as reduced availability of highly potent scaffolds or limitations of the Random Forest model, may also play a role.

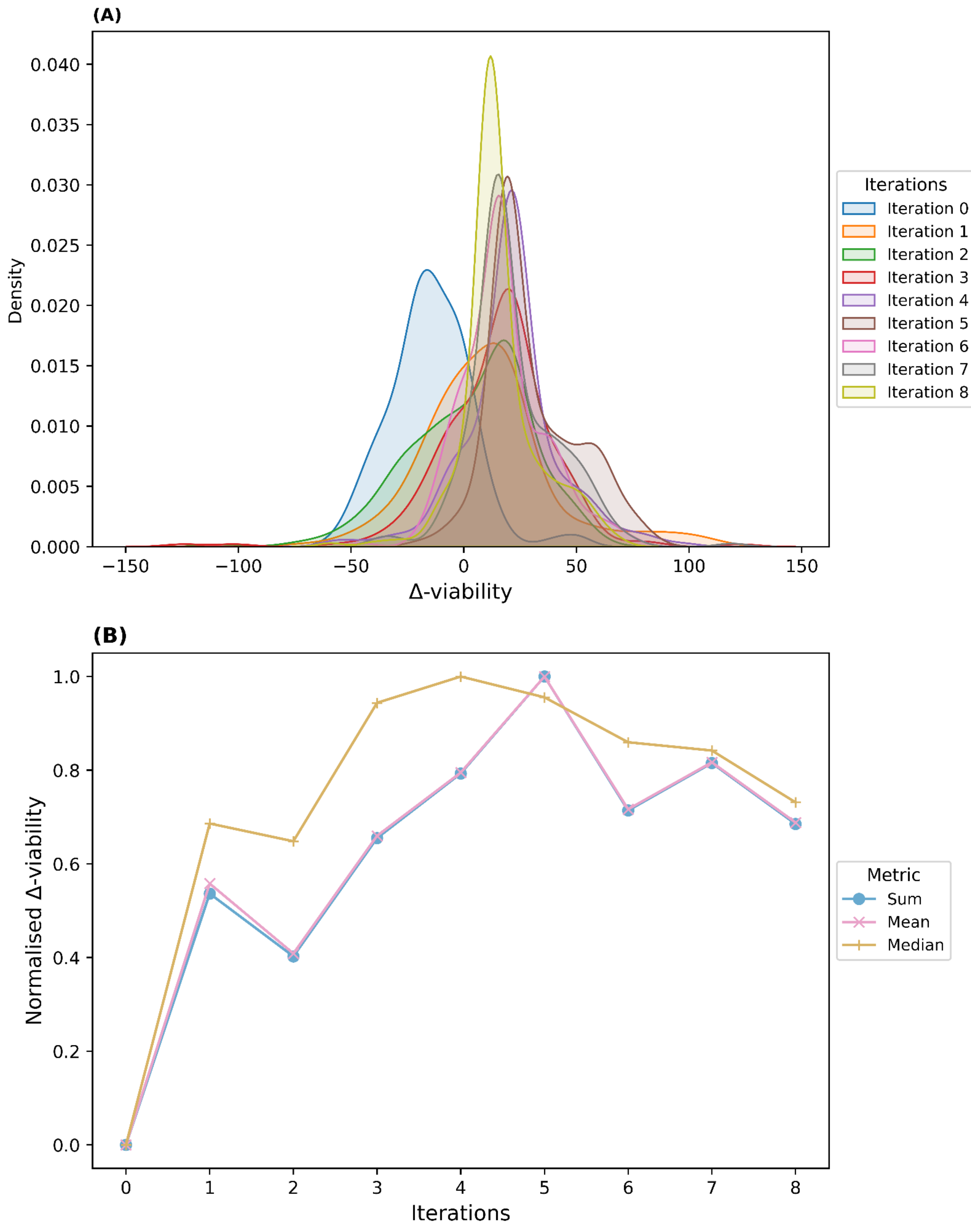

***Figure 1. Δ-viability distributions and iteration performance.*** *(A) The standard-scale Δ-viability distribution across iterative optimisation cycles, highlighting enrichment of highly selective combinations. Log-transformed distributions are provided in Fig. S2. (B) Difference in viability ($Viability_{MCF10A}$-$Viability_{MCF7}$) across eight iterations. The sum, mean and median of these differences are shown after normalisation.*

To evaluate predictive performance across optimisation iterations, we assessed cross-validated ranking performance using Spearman correlation between predicted and experimentally measured Δ-viability values. Correlation increased from 0.37 in the first model-guided iteration to a maximum of 0.59 by iteration 5, after which values fluctuated between 0.35 and 0.45 (Table S1). The peak in predictive performance coincided with the observed plateau in optimisation outcomes, suggesting that the model rapidly learned informative structure–activity relationships during early iterations and subsequently refined predictions within a narrower high-performing region of the search space.

Applying specific thresholds clarifies how optimisation improved selectivity (Fig. 2A-C). In the initial screen, ~30% of combinations reduced MCF7 viability below 30%, increasing to ~80% of combinations by iteration 6. Sparing MCF10A proved more challenging: only 14% of combinations maintained >50% viability in iteration 0, rising to ~30% by iteration 6. The number of combinations strongly impacting MCF7 increased steadily until approximately iteration 5 before stabilising. In contrast, MCF10A viability declined in early optimisation rounds as highly cytotoxic combinations were prioritised, before gradually recovering in later iterations. Early optimisation primarily enhanced tumour toxicity, with later rounds improving selectivity toward non-malignant cells. Tracking the most desirable combinations (MCF7 <30% and MCF10A >70%) reveals a fluctuating early trajectory followed by a consistent increase from iteration 3 onward. Across all iterations, 62 such combinations were identified for validation, of which 15 were confirmed.

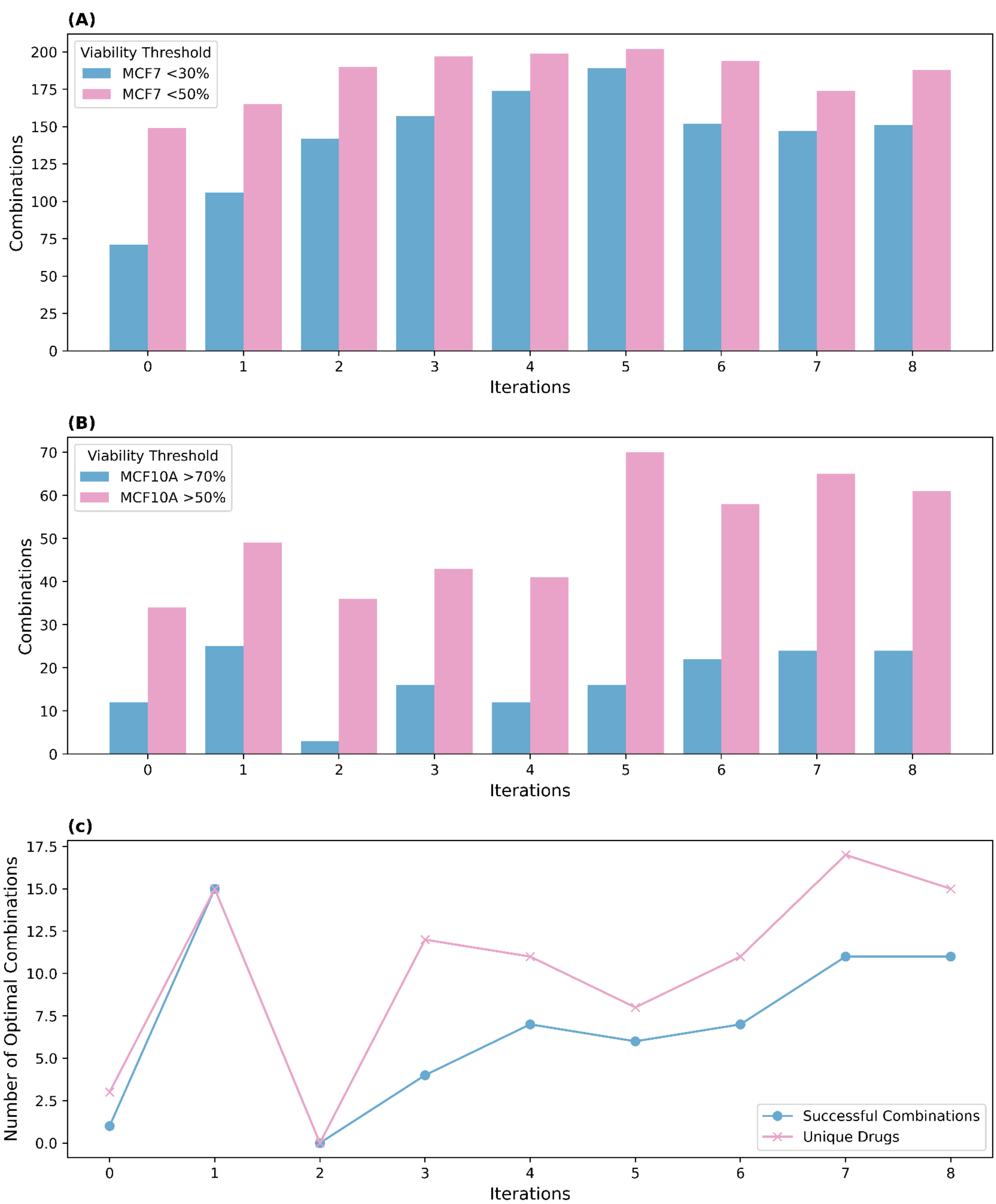

***Figure 2. Selectivity evolution for MCF7 and MCF10A.*** *Panels A–C illustrate how optimisation improved tumour-selective activity over successive iterations. Number of combinations toxic against MCF7 cells (A), and sparing of MCF10A (B). The number of identified combinations both showing high toxicity towards MCF7 (<30% viability) and sparing MCF10A (> 70% viability), as well as the total number of unique drugs that are part of these combinations are also indicated (C).*

## Chemical diversity of selected compounds

To determine whether optimisation occurred within a restricted chemical space or reflected broader exploration, we analysed the top 240 validated combinations (median, n=3) in which MCF10A viability exceeded 50%, ranked by differential viability favouring MCF7 toxicity. An important consideration is whether the algorithm converges on a narrow region of chemical space or continues to explore distinct areas. For translational relevance, it is essential that optimisation moves beyond incremental variations of a small subset of compounds.

Assessment of chemical diversity showed that validated hits span a broad region of the chemical space, indicating substantial exploration by the algorithm, while exhibiting a relative enrichment toward a leftward region (Fig. 3A). Ideal combinations (>70% viability for MCF10A and <30% viability for MCF7) were predominantly located within this region, with only rare outliers (Fig. 3B). In contrast, combinations exceeding a Δ-viability of 50% were more widely distributed, extending into central and right regions while still showing a bias toward the left (Fig. 3C). This distribution suggests that, although effective combinations are more frequently found within a specific subspace, the algorithm maintains coverage across a diverse chemical landscape rather than converging on a narrow set of closely related compounds. While the plots display all tested combinations, only validated hits (median, n=3) are highlighted.

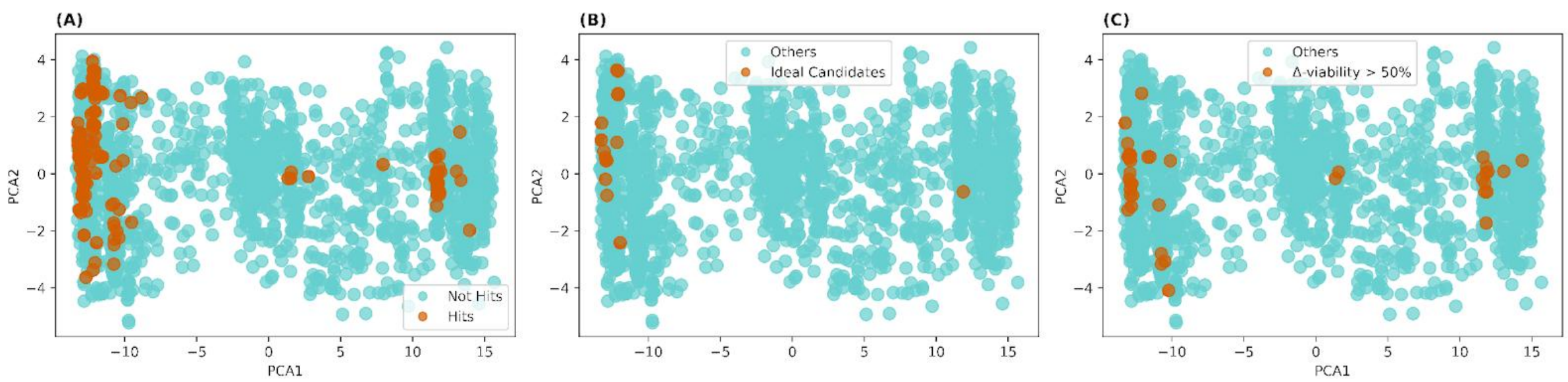

***Figure 3. Clustering of validated hits in combination chemical space.*** *All tested drug combinations are plotted based on their MACCS combination fingerprints. A) Validated hits (delta viability > 30% and MCF10A viability > 50%) are highlighted (n=147). B) Ideal candidates, MCF7 viability <30% and MCF10A viability >70% (n=15). C) Combinations with Δ-viability ≥50% and higher impact on MCF7 are also indicated (n=43). Triplet combinations were encoded using concatenated MACCS fingerprints. To remove ordering effects, fingerprints from all permutations of each triplet were averaged element-wise prior to PCA.*

**Effective compounds and their optimised combinations over iterations**

To evaluate whether potent compounds were further refined in successive optimisation cycles, we tracked two frequently selected agents, fulvestrant and vorinostat. We examined how the most effective combination (in terms of Δ-viability) containing each compound performed across iterations (Fig. 4). Δ-viability for combinations incorporating these drugs was rapidly maximised, with only modest gains in subsequent rounds. This early plateau suggests that high-value backbone agents were identified quickly, after which optimisation primarily refined partner selection and selectivity rather than further increasing maximal differential activity.

Analysis of individual drug contributions indicates that many of these combinations are synergistic and mechanistically diverse, targeting multiple molecular pathways **(Table S2)**. Together, these findings support a model in which the algorithm balances exploitation of strong core agents with continued diversification of mechanistic partners, rather than repeatedly converging on a single narrow solution space.

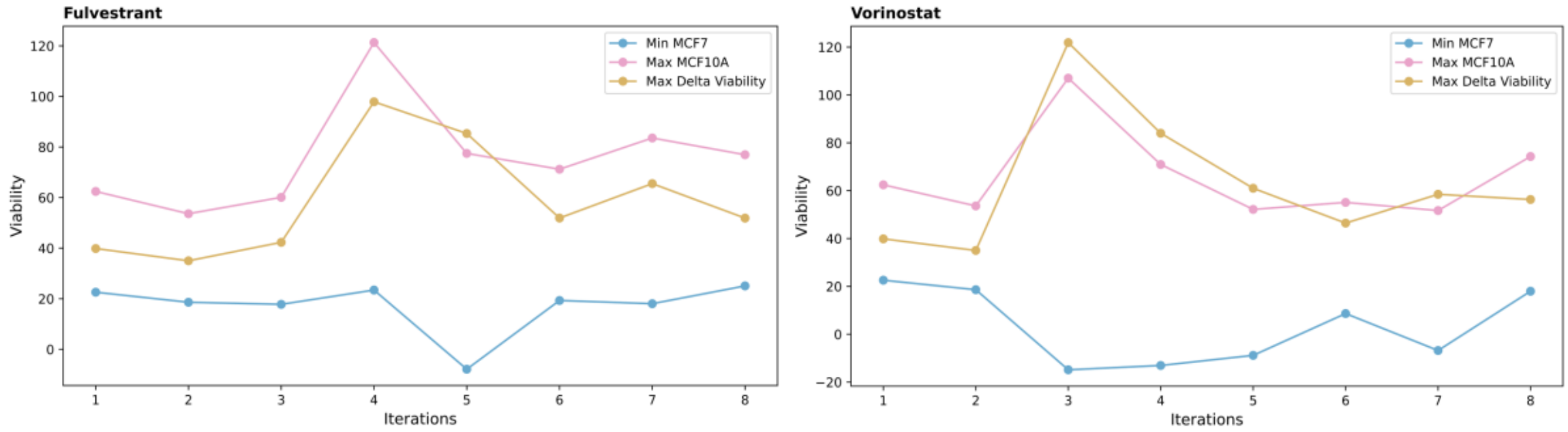

***Figure 4. Iteration tracking of key compounds (fulvestrant and vorinostat).*** *The top-ranked combination containing the indicated drug is shown across iterations, based on one of three criteria: minimum viability in MCF7 (blue), maximum viability in MCF10A (pink), or maximal Δ-viability (yellow). The corresponding Δ-viability values for these top-performing combinations are plotted on the y-axis.*

**Potent and selective drugs and synergistic combinations.**

For frequently selected compounds, $IC_{50}$ values were determined from independent triplicate (n = 3) dose–response experiments to assess intrinsic potency and confirm selective activity **(Table S3)**. The algorithm identified compounds spanning diverse chemical classes and molecular targets, indicating that optimisation did not converge on a single mechanistic axis.

Two representative combinations were examined in detail (Table 1a,b). Synergy was assessed using the Highest Single Agent (HSA) principle, where a combination is considered synergistic if its observed viability is lower than that of the most potent single agent at the same tested concentrations. In both cases, the triple-drug regimens produced greater toxicity in MCF7 cells than any individual constituent alone, while maintaining comparatively reduced effects on MCF10A, consistent with synergistic and selective activity.

| drug | $IC_{50_MCF7}$ (μM) | $IC_{50_MCF10}$ (μM) | administered conc. (μM) | viability MCF7 | viability MCF10 |
|---|---|---|---|---|---|
| Oxyphenisatin acetate | 0.11 | 10.48 | 0.155 | 39.72 | 85.65 |

| AZD5582 | 12.45 | 6.22 | 0.138 | 34.56 | 89.75 |
|---|---|---|---|---|---|
| Pevonedistat | 0.06 | 4.00 | 0.353 | 37.62 | 65.67 |
| Combination | | | | **17.24** | **70.32** |

*Table 1a. Combination 1. Calculated $IC_{50}$ values from three independent biological replicates (n = 3), together with resultant viability in MCF7 and MCF10A at the tested concentration.*

| drug | $IC_{50_MCF7}$ (µM) | $IC_{50_MCF10}$ (µM) | administered conc. (µM) | viability MCF7 | viability MCF10 |
|---|---|---|---|---|---|
| Fulvestrant | 22.074 | - | 0.114 | 39.72 | 74.25 |
| Pictilisib | 0.066 | 0.24 | 0.15 | 34.56 | 57.08 |
| buparlisib | 0.082 | 0.34 | 0.23 | 37.62 | 61.00 |
| Combination | | | | **26.49** | **73.33** |

*Table 1b. Combination 2. Calculated $IC_{50}$ values from three independent biological replicates (n = 3), together with resultant viability in MCF7 and MCF10A at the tested concentration.*

The mechanistic interpretation of these combinations aligns with known biology. In Combination 1 sensitivity to oxyphenisatin acetate, a protein synthesis inhibitor, has been reported in ER-positive cell lines including MCF7 [19]. Reduced responsiveness in MCF10A may reflect relatively higher CDKN1A expression, potentially limiting apoptosis induction, while elevated BIRC2/3 and CFLAR could further dampen sensitivity to SMAC mimetics such as AZD5582 by restricting caspase-8 activation [20].

In Combination 2 MCF7 sensitivity to Fulvestrant is consistent with elevated ESR1 expression, while concurrent inhibition of the PI3K pathway by pictilisib and buparlisib aligns with activation of downstream signalling components such as AKT1 and RPS6KB1 [21]. This dual targeting of hormone signalling and growth pathways is consistent with previously reported sensitisation of MCF7 cells to endocrine therapy through PI3K inhibition [22], supporting the observed combination effect.

To assess whether these mechanisms reflect broader trends rather than combination-specific effects we examined single agents with greater impact in MCF7 than MCF10A (Table 2). Fulvestrant targets ESR1, which is absent in MCF10A, this is consistent with tumour-selective activity. Foretinib may exploit RET upregulation in MCF7, implicating MAPK pathway signalling [23]. Vorinostat may show enhanced efficacy in MCF7 due to elevated CDKN1A expression, increasing susceptibility to HDAC-mediated cell-cycle arrest. Buparlisib targets PI3K signalling, consistent with increased downstream mTOR pathway activation in MCF7. Finally, pevonedistat activity is consistent with higher NEDD8 expression in MCF7 relative to MCF10A, suggesting increased reliance on neddylation.

While these hypotheses require further experimental validation, they offer biologically plausible explanations for the differential responses observed. Examination of the implicated targets reveals involvement of diverse cellular processes, including cell cycle regulation, apoptosis, growth signalling, and stress-response pathways. This mechanistic breadth underscores the capacity of the optimisation framework to identify combinations acting across multiple biological axes, an important consideration for limiting adaptive resistance in cancer therapy.

| drug | targets | Hypothesis for MCF7 sensitivity |
|---|---|---|
| fulvestrant | Estrogen receptor alpha (ERα) | MCF-7 highly expresses ERα, which fulvestrant degrades, leading to DNA damage and apoptosis. MCF-10A lacks ERα and is unaffected |
| foretinib | RET | MCF7 cells overexpress RET and key MAPK pathway components (GRB2, MAPK3), making them dependent on RET-mediated MAPK signaling. Foretinib inhibits RET, thereby selectively impairing MCF7 viability via disruption of this critical signaling axis. In contrast, RET expression levels in MCF10A are lower. |
| vorinostat | HDAC | MCF7 cells exhibit elevated baseline expression of CDKN1A, indicating a primed state for cell cycle arrest upon HDAC inhibition by vorinostat, which underlie their sensitivity. MCF10A may have more robust DNA repair or survival signaling pathways that buffer vorinostat's effects. |
| buparlisib | Pan-PI3K | MCF7 cells overexpress RPS6KB1, indicating elevated mTOR signaling downstream of PI3K, which makes them more vulnerable to PI3K inhibition by buparlisib. The RPS6KB1 levels are lower in MCF10A. |
| pevonedistat | NEDD8-activating enzyme (NAE) | MCF7 cells exhibit elevated NEDD8 expression compared to non-cancerous MCF10A, suggesting increased reliance on neddylation, which may underlie their heightened sensitivity to pevonedistat. |

*Table 2. Examples of single drug treatments that were more effective in MCF7 than MCF10A. Their main targets and a hypothesis for the differential response in the two cell lines are also provided.*

**Discussion**

Cancer therapy is constrained by the combinatorial complexity of multi-agent regimens and the heterogeneity of tumour responses. Traditional approaches, relying on static screening or retrospective analysis, cannot efficiently explore selective higher-order drug combinations across millions of possibilities. Here, we present a proof-of-concept framework in which iterative AI-guided modelling is combined with experimental testing to prioritise selective triple-drug therapies, demonstrating how machine learning can guide exploration of vast combinatorial spaces while retaining experimental oversight. While active learning has been applied previously to pairwise combination screens, few studies have addressed higher-order combinations with explicit multi-objective optimisation and iterative experimental refinement. By moving beyond pairwise interactions toward selective, multi-agent regimens, this approach highlights a scalable strategy for next-generation precision oncology and demonstrates the potential of iterative AI-guided workflows to accelerate rational therapeutic design.

Our results show that the framework efficiently identifies potent and selective drug combinations capable of distinguishing between tumorigenic MCF7 and non-tumorigenic MCF10A cells. The highest-ranking combinations substantially reduced MCF7 viability while maintaining MCF10A viability above 70%, and exhibited greater efficacy than their individual components, consistent with synergistic interactions. By explicitly optimising differential viability, the system prioritised regimens that balance tumour cytotoxicity with preservation of healthy cells, an essential objective in combination therapy design. Importantly, this represents a formal multi-objective optimisation strategy, simultaneously maximising efficacy and selectivity, rather than focusing solely on tumour cell killing.

The iterative active-learning cycle promoted both efficiency and breadth of exploration. Selected combinations spanned chemically distinct regions and involved compounds targeting diverse biological pathways, indicating that optimisation did not converge on a single mechanistic axis. Constraints limiting repeated use of individual compounds likely helped maintain this diversity. Despite the inherent risk of local optima, the sustained chemical and mechanistic spread suggests the search space remained broad, and experimental validation confirmed that predicted synergies were biologically meaningful as well as statistically robust.

Conceptually, our presented framework differs from conventional combination screening approaches by treating therapeutic discovery as an optimisation problem, rather than a fixed screening campaign. Instead of exhaustively testing predefined drug combinations, the system iteratively navigates a vast combinatorial space by learning from experimental outcomes and prioritising promising regions for subsequent exploration. Given the approximately 13.7 million possible triple-drug combinations in the screened library, identifying highly selective regimens from a comparatively small number of experiments illustrates the practical scalability of this optimisation strategy.

More broadly, this work supports the integration of AI with automated experimentation as a scalable strategy for next-generation therapeutic discovery. Beyond patient-specific applications, this framework can systematically identify optimised drug combinations for defined cancer subtypes or molecular contexts, enabling discovery of highly synergistic regimens at the population level. While this study involved human oversight during iterative experimentation, the workflow demonstrates the potential for adaptive, feedback-informed optimisation, efficiently prioritising promising combinations from a vast search space and refining model predictions based on experimental outcomes.

Several limitations should be acknowledged. This study establishes the feasibility of closed-loop, machine-learning-guided optimisation of multi-drug combinations using a single cancerous and non-tumorigenic cell line pair. Future work should extend the framework to additional cancer models and larger compound libraries to assess generalisability and translational potential. In this proof-of-concept, selection prioritised combinations with the highest predicted Δ-viability; to prevent overrepresentation of highly active compounds constraints limited the frequency of any single agent and allowed only one concentration per drug, maintaining chemical diversity and optimisation efficiency. Alternative strategies, such as uncertainty- or diversity-guided selection, could further balance exploration and exploitation. While our mechanistic interpretations are biologically plausible, additional experimental validation will be needed to establish causal links between implicated pathways and observed selective responses.

In summary, we present a proof-of-concept for autonomous, machine-learning-guided optimisation of triple-drug combinations, achieving selective and mechanistically diverse regimens against breast cancer cells. By integrating AI-driven modelling, automated experimentation, and experimental validation within a single adaptive framework, this study outlines a scalable approach to iterative therapeutic discovery in precision medicine.

## Methods

### Library

We used a library of 188 small-molecule drugs (see Appendix A). Putative $IC_{50}$ values were collected from the literature, and $IC_{10}$, $IC_{20}$, and $IC_{30}$ concentrations were extrapolated using the Hill equation (slope = 1). These concentrations were used to account for dose effects in combination experiments.

### Delta ($\Delta$) viability

Δ-viability was defined as a measure of tumor selectivity, representing a combination's ability to kill MCF7 breast cancer cells while sparing non-tumorigenic MCF10A cells. It was calculated by subtracting the viability of MCF7 from that of MCF10A. In an ideal case, a Δ-viability of 100 corresponds to complete killing of MCF7 cells with no effect on MCF10A.

### Iterative workflow

The workflow began with an initial randomly selected cohort of 230 combinations, designated as Iteration 0. In all subsequent optimisation iterations (1–8), the top 230 predicted combinations were selected by the model. To monitor assay consistency, five combinations from the previous iteration were randomly added to these top predictions, resulting in 235 combinations tested per iteration. When these random combinations were selected, only the new 230 model-predicted combinations were considered for ranking and selection.

Each iteration had a defined start and end: plate allocation of selected combinations, experimental testing, integration of data from all previous iterations, retraining of the predictive model, and prediction of all untested combinations. Iteration 0 was an exception, as all tested combinations were randomly selected rather than model-guided.

### Screening compounds

Cell Culture: Cell viability was measured for MCF7 and MCF10A cell lines. Cells were cultured in their respective growth media at 37°C in a humidified atmosphere containing 5% $CO_2$. Media were replaced every 2–3 days, and subculturing was performed at approximately 80% confluency (typically every 3–4 days) using 1X TrypLE Select for MCF7 and the PromoCell Detach Kit for MCF10A cells.

Compound Plate Preparation: Vehicle (DMSO) or designated compound mixtures were dispensed into white 384-well plates (one plate per cell line) in randomized layouts using a D300e digital dispenser. Each well volume was standardized to 404 nL, maintaining a final DMSO concentration of 0.5% (v/v) across the assay. Plates were sealed and stored at −4°C until required.

Viability Assay: Cells were seeded into prepared compound plates using a Dragonfly 10-Channel Dispenser at $5.0 \times 10^3$ cells/cm$^2$ (approximately 300 cells per well) in 40 µL of the appropriate medium. Plates were incubated for 48 hours at 37°C and 5% $CO_2$. After incubation, plates were equilibrated to 30°C, and 40 μL of CellTiter-Glo® 3D reagent (Promega, Cat. #G9682) was added. Plates were briefly centrifuged and incubated for 30 minutes in the dark at 30°C before luminescence was read using a CLARIOstar Plus microplate reader (BMG Labtech).

Data Analysis: Raw luminescence data (RLU) were derandomized and background-subtracted. Cell viability was expressed as a percentage relative to the vehicle control using the formula:
% Viability = (RLU – Mean Positive Control) / (Mean Negative Control – Mean Positive Control) × 100.

Quality control: In each iteration (from iteration 1-8), 5 combinations were re-tested from the previous iteration to ensure reasonable consistency in measurements across experiments (Table S4). A final validation of results was done for top combinations (Table S5), and single drug measurements most featured in the top combinations were validated (Table S3). The results showed a high degree of accuracy and reproducibility.

**Drug representation and combination generation:**
Each drug was represented using the Molecular ACCess System (MACCS) 2-dimensional fingerprint, consisting of 166 binary features using RDKit (https://www.rdkit.org). MACCS fingerprints were selected because they provide a compact representation suitable for early-iteration modelling with limited training data, reducing overfitting risk compared with longer fingerprints such as Morgan.

Each drug could be represented at one of three concentrations corresponding to its $IC_{10}$, $IC_{20}$, or $IC_{30}$ values. The selected concentration was appended to the MACCS vector, producing three possible representations per drug. Triple-drug combinations were generated from these drug representations, resulting in a 501-length vector per combination (three 167-length vectors concatenated). Each combination was assigned a unique identifier using the first 20 characters of its SHA-1 hash. Only one concentration representation per drug could appear in any given combination.

To avoid artefactual toxicity caused by high total chemical concentration, only combinations with a cumulative concentration below 50 μM were considered. This filtering produced 13,725,749 valid unique triple-drug combinations for modelling and experimental testing.

**Modelling, prediction, and selection**
For each iteration, a model was trained to predict Δ-viability for the set of unique triple-drug combinations. Because the order of drugs in a combination vector is not invariant and can influence what the model learns, each unique combination was represented by six samples corresponding to all possible order permutations. All six permutations shared the same combination identifier and Δ-viability value. Once a combination was experimentally tested, all its permutations were considered validated and removed from subsequent predictions. Data from all previous iterations were included in the model to improve prediction accuracy.

In the prediction phase, Δ-viability was predicted for all untested combinations. Combinations were ranked from highest to lowest predicted Δ-viability, and the top 235 unique combinations were selected for experimental testing. Two constraints were applied during selection: (1) each combination had to contain a unique set of three drugs, irrespective of dosage; if multiple dosage variations existed, the predicted best-performing combination was chosen, and (2) no single drug could appear in more than 20% of the selected combinations, limiting overrepresentation of highly active drugs.

Random Forests (R implementation; https://www.r-project.org) were used as the predictive algorithm due to their robustness to small sample sizes, particularly in early iterations. The model used 200 trees for iterations 0–2 and 1,000 trees from iteration 3 onwards, with all other parameters set to default. A complete model-based exploitation strategy was employed, selecting combinations solely based on predicted Δ-viability. This approach was chosen as a baseline for triple-drug combination optimisation; the benefits of balancing exploitation and exploration in this context remain to be explored.

**Comparative analysis**

Gene expression differences between MCF7 and MCF10A cells were assessed using publicly available transcriptomic data from the Gene Expression Omnibus (accession number GSE71862): https://www.ncbi.nlm.nih.gov/geo/geo2r/?acc=GSE71862

Drug target annotations, including primary and secondary targets, were obtained from curated databases including DrugBank and PubChem.

**Acknowledgements**

None of the above work would have been possible without the availability of cancer derived cell lines. We'd like to express gratitude to all patients who have made this, as well as many other studies contributing to the understanding and treatment of cancer, a possibility. We'd also like to thank all contributors to the tools we've used in the analysis of this study.

This work has been supported by grants from the UK Engineering and Physical Sciences Research Council (EPSRC) [EP/R022925/2, EP/W004801/1 and EP/X032418/1].

**Data and code availability**

All data and code has been made available through a github repository: https://github.com/oghenejokpeme/multidrug-analysis

**Competing interests**

R.C., E.B., E.T., and H.S. works with Arctoris ltd, a CRO in life sciences. The remaining authors declare no competing interests.

**References**

1. Cherkasov, A., Muratov, E. N., Fourches, D., Varnek, A., Baskin, I. I., Cronin, M., Dearden, J., Gramatica, P., Martin, Y. C., Todeschini, R., Consonni, V., Kuz'min, V. E., Cramer, R., Benigni, R., Yang, C., Rathman, J., Terfloth, L., Gasteiger, J., Richard, A., & Tropsha, A. (2014). QSAR modeling: where have you been? Where are you going to?. *Journal of medicinal chemistry*, *57*(12), 4977–5010. https://doi.org/10.1021/jm4004285

2. Al-Lazikani, B., Banerji, U., & Workman, P. (2012). Combinatorial drug therapy for cancer in the post-genomic era. *Nature biotechnology*, *30*(7), 679–692.

https://doi.org/10.1038/nbt.2284

3. Vamathevan, J., Clark, D., Czodrowski, P., Dunham, I., Ferran, E., Lee, G., Li, B., Madabhushi, A., Shah, P., Spitzer, M., & Zhao, S. (2019). Applications of machine learning in drug discovery and development. *Nature reviews. Drug discovery*, *18*(6), 463–477. https://doi.org/10.1038/s41573-019-0024-5

4. Tropsha, A., Isayev, O., Varnek, A., Schneider, G., & Cherkasov, A. (2024). Integrating QSAR modelling and deep learning in drug discovery: the emergence of deep QSAR. *Nature reviews. Drug discovery*, *23*(2), 141–155. https://doi.org/10.1038/s41573-023-00832-0

5. Letai A. (2017). Functional precision cancer medicine-moving beyond pure genomics. Nature medicine, 23(9), 1028–1035. https://doi.org/10.1038/nm.4389

6. Prasad, V. Perspective: The precision-oncology illusion. Nature 537, S63 (2016). https://doi.org/10.1038/537S63a

7. Abdel-Rehim, A., Orhobor, O., Griffiths, G., Soldatova, L., & King, R. D. (2025). Establishing predictive machine learning models for drug responses in patient derived cell culture. *NPJ precision oncology*, *9*(1), 180. https://doi.org/10.1038/s41698-025-00937-2

8. Kornauth, C. et al., (2022). Functional Precision Medicine Provides Clinical Benefit in Advanced Aggressive Hematologic Cancers and Identifies Exceptional Responders. Cancer discovery, 12(2), 372–387. https://doi.org/10.1158/2159-8290.CD-21-0538

9. Dolgin E. (2024). The future of precision cancer therapy might be to try everything. Nature, 626(7999), 470–473. https://doi.org/10.1038/d41586-024-00392-2

10. Dobrolecki, L. E. et al., (2016). Patient-derived xenograft (PDX) models in basic and translational breast cancer research. Cancer metastasis reviews, 35(4), 547–573. https://doi.org/10.1007/s10555-016-9653-x

11. Weeber, F. et al., (2015). Preserved genetic diversity in organoids cultured from biopsies of human colorectal cancer metastases. Proceedings of the National Academy of Sciences of the United States of America, 112(43), 13308–13311. https://doi.org/10.1073/pnas.1516689112

12. Wensink, G.E., Elias, S.G., Mullenders, J. et al. Patient-derived organoids as a predictive biomarker for treatment response in cancer patients. npj Precis. Onc. 5, 30 (2021). https://doi.org/10.1038/s41698-021-00168-1

13. Liebers, N. et al., Ex vivo drug response profiling for response and outcome prediction in hematologic malignancies: the prospective non-interventional SMARTrial. Nat Cancer 4, 1648–1659 (2023). https://doi.org/10.1038/s43018-023-00645-5

14. Snijder, B. et al., (2017). Image-based ex-vivo drug screening for patients with aggressive haematological malignancies: interim results from a single-arm, open-label, pilot study. The

Lancet. Haematology, 4(12), e595–e606. https://doi.org/10.1016/S2352-3026(17)30208-9

15. Kornauth, C. et al., (2022). Functional Precision Medicine Provides Clinical Benefit in Advanced Aggressive Hematologic Cancers and Identifies Exceptional Responders. Cancer discovery, 12(2), 372–387. https://doi.org/10.1158/2159-8290.CD-21-0538

16. Wang, S., Allauzen, A., Nghe, P., & Opuu, V. (2025). A guide for active learning in synergistic drug discovery. *Scientific reports*, *15*(1), 3484. https://doi.org/10.1038/s41598-025-85600-3

17. Tosh, C., Tec, M., White, J. B., Quinn, J. F., Ibanez Sanchez, G., Calder, P., Kung, A. L., Dela Cruz, F. S., & Tansey, W. (2025). A Bayesian active learning platform for scalable combination drug screens. *Nature communications*, *16*(1), 156. https://doi.org/10.1038/s41467-024-55287-7

18. Jin, S., Li, X., Yang, G., Zhang, Z., Shi, J. Q., Liu, Y., & Zhao, C. X. (2025). Active Learning-Based Prediction of Drug Combination Efficacy. *ACS nano*, *19*(18), 17929–17940. https://doi.org/10.1021/acsnano.5c04810

19. Morrison, B. L., Mullendore, M. E., Stockwin, L. H., Borgel, S., Hollingshead, M. G., & Newton, D. L. (2013). Oxyphenisatin acetate (NSC 59687) triggers a cell starvation response leading to autophagy, mitochondrial dysfunction, and autocrine TNFα-mediated apoptosis. *Cancer medicine*, *2*(5), 687–700. https://doi.org/10.1002/cam4.107

20. Cong, H., Xu, L., Wu, Y., Qu, Z., Bian, T., Zhang, W., Xing, C., & Zhuang, C. (2019). Inhibitor of Apoptosis Protein (IAP) Antagonists in Anticancer Agent Discovery: Current Status and Perspectives. *Journal of medicinal chemistry*, *62*(12), 5750–5772. https://doi.org/10.1021/acs.jmedchem.8b01668

21. Zhang, H. P., Jiang, R. Y., Zhu, J. Y., Sun, K. N., Huang, Y., Zhou, H. H., Zheng, Y. B., & Wang, X. J. (2024). PI3K/AKT/mTOR signaling pathway: an important driver and therapeutic target in triple-negative breast cancer. *Breast cancer (Tokyo, Japan)*, *31*(4), 539–551. https://doi.org/10.1007/s12282-024-01567-5

22. Campone, M., Im, S. A., Iwata, H., Clemons, M., Ito, Y., Awada, A., Chia, S., Jagiełło-Gruszfeld, A., Pistilli, B., Tseng, L. M., Hurvitz, S., Masuda, N., Cortés, J., De Laurentiis, M., Arteaga, C. L., Jiang, Z., Jonat, W., Le Mouhaër, S., Sankaran, B., Bourdeau, L., … Baselga, J. (2018). Buparlisib plus fulvestrant versus placebo plus fulvestrant for postmenopausal, hormone receptor-positive, human epidermal growth factor receptor 2-negative, advanced breast cancer: Overall survival results from BELLE-2. *European journal of cancer (Oxford, England : 1990)*, *103*, 147–154. https://doi.org/10.1016/j.ejca.2018.08.002

23. Regua, A. T., Najjar, M., & Lo, H. W. (2022). RET signaling pathway and RET inhibitors in human cancer. *Frontiers in oncology*, *12*, 932353. https://doi.org/10.3389/fonc.2022.932353